\documentclass[a4paper,11pt]{article}
\usepackage{cite}
\pdfoutput=1
\usepackage{amsmath}
\usepackage{graphicx}
\usepackage{amsfonts}
\usepackage{amssymb}
\usepackage{amsmath}
\usepackage{amsthm}
\usepackage{amscd}
\usepackage{dsfont}
\usepackage{mathtools}
\usepackage{bm}
\usepackage[svgnames]{xcolor}
\usepackage[colorlinks=true,urlcolor=purple,linkcolor=violet,citecolor=magenta,bookmarks=false]{hyperref}
\usepackage{stmaryrd}
\usepackage[nomessages]{fp}
\usepackage{color}
\usepackage{hyperref}
\hypersetup{colorlinks,linkcolor={blue},citecolor={blue},urlcolor={blue}}
\usepackage{empheq}

\definecolor{orange-lpens}{RGB}{230,62,37}
\definecolor{bleu-lpens}{RGB}{41,50,117}

\numberwithin{equation}{section}
\newcommand{\bc}{\begin{center}}
\newcommand{\ec}{\end{center}}
\newcommand{\beq}{\begin{equation}}
\newcommand{\eeq}{\end{equation}}
\newcommand{\beqa}{\begin{eqnarray}}
\newcommand{\eeqa}{\end{eqnarray}}
\newcommand{\beqs}{\begin{eqnarray*}}
\newcommand{\eeqs}{\end{eqnarray*}}

\newcommand{\bi}{\begin{itemize}}
\newcommand{\ei}{\end{itemize}}

\def\vev#1{\langle #1\rangle}
\def\cadremath#1{\vbox{\hrule\hbox{\vrule\kern8pt\vbox{\kern8pt
			\hbox{ {$\displaystyle #1 $ } }\kern8pt} 
			\kern8pt\vrule}\hrule}}
\theoremstyle{definition}

\theoremstyle{definition}

\theoremstyle{definition}

\theoremstyle{definition}

\newcommand{\EE}{ \mathbb{E} }
\newcommand{\DD}{ \mathbb{D} }

\newcommand{\Ff}{ \mathcal{F} }
\newcommand{\Oo}{ \mathcal{O} }

\newcommand{\pp}{{\mathsf{p}}}
\newcommand{\qq}{{\mathsf{q}}}

\begin{document}

\begin{center}
{\Large \bf  The Renormalization Group as a Stochastic Exploration Process }
\end{center}

\vspace{0.3cm}

\begin{center}
Denis BERNARD \footnote{Email: denis.bernard@phys.ens.fr}
\end{center}

\noindent
{\it Laboratoire de Physique de l'\'Ecole Normale Sup\'erieure, CNRS, ENS \& Universit\'e PSL, Sorbonne Universit\'e, Universit\'e Paris Cit\'e, 75005 Paris, France.}
\vspace{0.3cm}

\centerline{\today}

\vspace{0.5cm}

\begin{abstract}
The Renormalization Group (RG) is a powerful and versatile framework for analyzing complex physical systems. Here, we reinterpret it as a stochastic process that explores physical phase spaces, scale by scale, progressively revealing finer details of small-scale structures. This perspective establishes a natural connection to other random exploration processes, such as the Schramm-Loewner evolution, and is more suited for a probabilist audience. It links RG concepts such as RG transformations, effective actions, etc, to usual probabilistic tools such as conditional expectation values, martingales, etc, but also makes contact with stochastic quantization.
\end{abstract}
\vspace{0.3cm}

{\setlength{\parskip}{0mm} \setlength{\baselineskip}{0.05 in} \tableofcontents}

\medskip
\centerline{-----------------------------}

\section{Introduction}
\label{sec:intro}

Complex physical systems involve numerous intertwined degrees of freedom from which emerge collective patterns~\cite{Anderson1972}. 
By organizing the system analysis scale by scale, the Renormalization Group (RG) provides a powerful framework to reveal the relevant collective degrees of freedom and their dynamics~\cite{Wilson1971-a,Wilson1971-b}, with numerous applications across diverse disciplines~\cite{RevModPhys.46.597,Kadanoff-book,amit1984field,sachdev2011quantum}.
For a pedagogical introduction to the RG and its applications, we refer the reader to the elegant treatment by J. Cardy~\cite{Cardy1996}. 

In this note, we draw an analogy between the RG and the Schramm-Loewner Evolution (SLE)~\cite{Schramm:2000,Lawler:2007sle} -- a stochastic process designed to describe random interfaces in two-dimensional critical systems.
Both SLE and RG are exploration processes, albeit organized by distinct principles: while SLE explores interfaces as a function of their length, RG explores physical phase spaces as a function of the length scales of field fluctuations.
In both cases, they allow to reshuffle the statistical sum to reveal the relevant degrees of freedom.

By framing the RG as a stochastic process, we aim to make its structure more accessible to probabilists. This perspective requires introducing filtrations of $\sigma$-algebras on phase spaces, which encode progressively finer detailed information on field configurations. The RG procedure of "integrating out" small-scale fluctuations then translates to computing conditional expectational values with respect to these $\sigma$-algebras. Consequently the RG flow emerges as a flow of conditional expectation values, as usual in stochastic processes.
The semi-group property of RG transformations arises naturally from the nested structure of conditional expectation values, while the martingale properties of conditional expectation values illuminate many natural RG structures -- bridging concepts from probability theory and the Renormalization Group.

We begin in Section \ref{sec:SLE} by illustrating the role of stochastic processes in statistical physics through SLE. In Section \ref{sec:RG-exploration}, we formalize the RG from an exploration perspective. Sections \ref{sec:blocks} and \ref{sec:gaussian} present two standard approaches to defining the RG exploration process, while Section \ref{sec:scaling} demonstrates how this viewpoint clarifies the general RG principles underlying  the definition of field theories in the continuum and their renormalization.
\medskip

{\it Added note:} This note is an expanded version of a lecture given by the author at the ''Peking summer program in probability: Probability \& Quantum Field Theory" in 2026. The possibility to view the RG as an exploration process, and its analogy with SLE, was understood by Michel Bauer and the author while  studying SLE and its correspondance with conformal field theory (CFT), in 2002-2006. See e.g. refs.\cite{Bauer2005Multiple,Bauer:2006growth} on SLE \& CFT, with comments about analogies with RG, or the author's lecture notes on statistical field theory~\cite{Bernard-SFTnotes}. In the meantime, a similar point of view was nicely and extensively developed by R. Bauerschmidt and T. Bodineau in~\cite{Bauerschmidt_2024}. This viewpoint also bares similarity with the approach to stochastic quantization advocated by I. Bailleul, I. Chevyrev and M. Gubinelli in~\cite{bailleul2023wilsonitodiffusions}.

\section{Conditioning and statistical martingales: the SLE case}
\label{sec:SLE}

In this Section we use the relation between interfaces in statistical physics and SLE \cite{Lawler:2007sle,Cardy:2005sle,Bauer:2006growth} to illustrate how stochastic processes emerge in statistical models and how they help studying them. The main point consists in associating those processes to partitioning -- or conditioning -- of the configuration spaces, and in organizing Boltzmann sums according to these partitioning.
We will discuss how to extend this viewpoint to the RG in Section \ref{sec:RG-exploration}.

\medskip

$\bullet$ \underline{\it Statistical interfaces and the domain Markov property.}
Let us first briefly recall the basic rules of statistical physics (at equilibrium). Imagine considering a finite domain $\DD$ in two dimensions, with the topology of a disk, filled with a lattice $\Gamma$. Some degrees of freedom, called spins, sit at each vertex of the lattice. A configuration $c$ is the data of all spin values, i.e. the data of all $s_{i}$ for $i\in\Gamma$. Let $\mathcal{C}$ be the set of configurations (which is finite for a discrete model in a finite volume).  A probability measure -- the Boltzmann measure -- is defined on $\mathcal{C}$ by assigning to each configuration $c\in\mathcal{C}$ the probability 
\beq
\mathbb{P}_{c} := Z_\mathbb{D}^{-1}\, w^\mathbb{D}_{c} ,
\eeq
with $w^\mathbb{D}_{c}$ the so-called Boltzmann weights, $w^\mathbb{D}_{c} = e^{- E^\mathbb{D}_{c}}$ with $E^\mathbb{D}_c$ some configuration 'energy'.
The normalization factor $Z_\mathbb{D} = \sum_{c\in C}w^\mathbb{D}_{c}$ is called the partition function.  
It depends on the shape of the domain $\DD$.

For the Ising model, the spins take value $\pm 1$, i.e. $s_{i} = \pm 1$ for $i\in\Gamma$, and the energy $E^\mathbb{D}_c$ is of the form $ E^\mathbb{D}_{c} =- \sum_{i\sim j} J s_{i}s_{j}$, where the sum is over neighboring sites. Given a spin configuration, some sites are assigned $+$, others $-$, and we aim at understanding the shape of the curves interfacing the clusters of given spin values.  Following~\cite{Schramm:2000}, let us fix two points $x_{0}$ and $x_{\infty}$ on the boundary, and assign $+$ on all boundary sites from $x_{0}$ to $x_{\infty}$ (going counterclockwise) and $-$ on the ones from $x_{\infty}$ to $x_{0}$. This ensures the existence of an interface from $x_{0}$ to $x_{\infty}$, for each configuration $c\in\mathcal{C}$. The Boltzmann measure then induces a probability measure on curves $\gamma$ from $x_0$ to $x_\infty$,
\beq \label{eq:P-gamma}
P_{\DD,x_{0},x_{\infty}}[\gamma] := \frac1{Z_{\DD}}\sum_{c\in \mathcal{C}[\gamma]}w^\mathbb{D}_{c} =\frac{Z_{\DD}[\gamma]}{Z_{\DD}}, 
\eeq
with $Z_{\DD}[\gamma]:= \sum_{c\in \mathcal{C}[\gamma]}w^\mathbb{D}_{c}$, where $\mathcal{C}[\gamma]$ is the subset of configurations giving $\gamma$ as the interface, i.e. the set of configurations such that the spins immediately to the left of $\gamma$ are $-$ and immediately to the right are $+$.

Although we introduced this measure using the Ising model, it can similarly be defined for a large class of lattice statistical models (percolation, self-avoiding model, Potts models, $O(n)$ model, etc). What the definition of SLE achieved~\cite{Schramm:2000} is to given a precise meaning to this measure in the continuum (when the lattice spacing vanishes) and at criticality (to make use of conformal geometry).

The measure  \eqref{eq:P-gamma}  is a measure on curves from $x_{0}$ to $x_{\infty}$.  It fulfills to a property, called the domain Markov property~\cite{Schramm:2000}, whenever the interaction between the spins is local enough.  Given a point $x$ in the bulk of $\DD$ and a piece of curve $\gamma_{x_{0},x}$ from $x_{0}$ to $x$ that we view as a portion of an interface, we can consider either the measure  \eqref{eq:P-gamma}  conditioned on $\gamma_{x_{0},x}$, or the measure  \eqref{eq:P-gamma} but in the domain cut along $\gamma_{x_{0},x}$. The domain Markov property claims that both are identical:
\beq \label{eq:domain-markov}
P_{\DD;x_{0},x_{\infty}}[\,\cdot\,|\gamma_{x_{0},x}] = P_{\DD \setminus \gamma_{x_{0},x};x,x_{\infty}}[\,\cdot\,] .
\eeq

Let us prove it starting from \eqref{eq:P-gamma}. The support of both measures in \eqref{eq:domain-markov} is identically made of curves from $x$ to $x_\infty$. Call $\hat\gamma$ the variable curve from $x$ to $x_{\infty}$.  By the definition of conditional probability, we have:
$$ 
P_{\DD;x_{0},x_{\infty}}[\gamma_{x_{0}x}\hat\gamma| \gamma_{x_{0}x}] = \frac{P_{\DD;x_{0},x_{\infty}}[\gamma_{x_{0}x}\hat\gamma]}{P_{\DD;x_{0},x_{\infty}}[\gamma_{xx_{0}}]} ,
$$
with $P_{\DD;x_{0},x_{\infty}}[\gamma_{xx_{0}}] = Z_{\DD}[\gamma_{xx_{0}}]/Z_{\DD}$.  The conditioned partition function $Z_{\DD}[\gamma_{xx_{0}}]$ involves the sum over all spins away from $\gamma$, so that $Z_{\DD;x_{0},x_{\infty}}[\gamma_{x_{0}x}] = e^{- E_{\gamma_{x_{0}x}}}\, Z_{\DD \setminus \gamma_{x_{0}x};x,x_{\infty}}$, where the factor ${E_{\gamma_{x_{0}x}}}$ is the energy along the curve $\gamma_{x_{0}x}$. This extra factor arises because the lattice edges crossing $\gamma_{x_{0}x}$ are not part of the Boltzmann weights in $\DD \setminus \gamma_{x_{0}x}$, while they are included in those defined in $\DD$. This factor cancels out when computing the conditional probability, and we get:
$$ 
\frac{P_{\DD}[\gamma_{x_{0}x}\hat\gamma]}{P_{\DD}[\gamma_{xx_{0}}]} = \frac{Z_{\DD}[\gamma_{x_{0}x}\hat\gamma]}{Z_{\DD}[\gamma_{xx_{0}}]} = \frac{Z_{\DD\setminus \gamma_{x_{0}x}}[\hat \gamma]}{Z_{\DD\setminus \gamma_{x_{0}x}}} = P_{\DD\setminus \gamma_{x_{0}x};x,x_{\infty}}[\hat\gamma],
$$
which proves \eqref{eq:domain-markov}. This property is valid at or away from criticality. 
\medskip

$\bullet$ \underline{\it Boltzmann sums, conditioning and statistical martingales.}
Following~\cite{Schramm:2000}, let us now associate the measure \eqref{eq:P-gamma} to an exploration process coding for the growth of the interfaces. This leads to formulate a relation between statistical mechanics and martingale processes, and to observe that martingales provide appropriate tools to efficiently organize Boltzmann statistical sums~\cite{Bauer_2003,Bauer2005Multiple}.

Given the domain $\DD$, and proper boundary conditions enforcing the presence of an interface from $x_0$ to $x_\infty$, we consider a portion of an interface $\gamma_{T}$ emerging from $x_{0}$ of fixed length ${T}$. Let $\mathcal{C}[\gamma_{T}]$ be the subset of configurations giving $\gamma_T$ as the interface, i.e. with $-$ $(+)$ spins on the left (right) of $\gamma_T$. The collection of $\mathcal{C}[\gamma_{T}]$ forms a partition of the total configuration space:
\beq
\mathcal{C} = \bigcup_{\gamma_{T}}\mathcal{C}[\gamma_{T}] .
\eeq
This partition becomes finer as the length $T$ increases. It thus defines a filtration $\Ff_{T}$ of $\sigma$-algebras over $\mathcal{C}$, such $\Ff_{T} \subseteq \Ff_{S}$ for $S>T$, that specifies our exploration process, with probability measure that induced by the Boltzmann rules.  

Any given observable $\Oo$ -- maybe you measure the values of spins at various locations -- assigns a value $\Oo_{c}$ for each configuration $c\in \mathcal{C}$, so that observables are functions on the configuration space. Their expectation values are defined by the Boltzmann sums:
\beq
\EE_{\DD}[\Oo]: = \frac1{Z_{\DD}} \sum_{c\in \mathcal{C}} w^\mathbb{D}_{c}\Oo_{c} .
\eeq

Given the above setup, we consider observable expectation values conditioned on a given portion $\gamma_T$ of the interface:
\beq
 \EE_\mathbb{D}[\Oo \vert \gamma_{T}]  = \frac1{Z_{\DD}[\gamma_{T}]} \sum_{c\in \mathcal{C}[\gamma_{T}]}w^\DD_{c}\Oo_{c} ,
 \eeq
with $Z_{\DD}[\gamma_{T}] = \sum_{c\in C[\gamma_{T}]} w_{c}^{\DD}$, as above.
As for the domain Markov property, we may also consider expectation values $\EE_{\DD\setminus\gamma_T}[\Oo]$ in the domain cut along the curve $\gamma_T$. 
\medskip

We then have the simple, but important remark~\cite{Bauer_2003,Bauer2005Multiple}. For any observable $\Oo$:\\
(i) $\EE_{\DD\setminus\gamma_T}[\Oo] = \EE_\mathbb{D}[\Oo \vert \gamma_{T}]$.\\
(ii) $T\to \EE_{\DD\setminus\gamma_T}[\Oo]$ is a martingale (w.r.t. the filtration $\mathcal{F}_T$).
\medskip

The first property follows  from the fact that the Boltzmann weights in the domain $\DD$ and in the cut domain $\DD\setminus \gamma_T$ only differ by a factor coding for the energy of the curve, $w_{c}^{\DD} = e^{-E_{\gamma_{T}}}w_{c}^{\DD\setminus\gamma_{T}}$, as discussed above. The second property follows from the first since $T\to \EE_\mathbb{D}[\Oo \vert \gamma_{T}]$ is a (tautological) martingale. 
Recall that a martingale is a $\Ff_{S}$-measurable map $S \mapsto M_{S}$ satisfying $\EE[M_{T}|\Ff_{S}] = M_{S}$ for $S<T$. This property holds for any map $M_S$ defined as a conditional expectation value w.r.t. $\Ff_{S}$, since $\EE\bigl[\EE[\Oo|\Ff_{T}]\big|\Ff_{S}\bigr] = \EE[\Oo|\Ff_{S}]$ for $S<T$. 
Let us illustrate this by explicitly checking that $\EE[\EE_{\DD\setminus\gamma_T}[\Oo]]=\EE_\mathbb{D}[\Oo]$. We have:
\begin{align}  \label{eq:Z-check}
\EE[\EE_{\DD\setminus\gamma_T}[\Oo]] &= \sum_{\gamma_{T}}\mathbb{P}_\mathbb{D}[\gamma_{T}] \EE_{\DD\setminus\gamma_T}[\Oo] \\
& = \sum_{\gamma_{T}} \frac{Z_{\DD}[\gamma_{T}]}{Z_{\DD}}\cdot  \frac1{Z_{\DD}[\gamma_{T}]} \sum_{c\in \mathcal{C}[\gamma_{T}]} w^\mathbb{D}_{c}\Oo_{c} = \frac1{Z_{\DD}} \sum_{c\in \mathcal{C}}w^\mathbb{D}_{c}\Oo_{c} = \EE_\mathbb{D}[\Oo] .
\nonumber
\end{align}
The martingale property is thus (almost) tautological in statistical physics. Checking it amounts to organize the Boltzmann sum according to the partitioning of the configuration space specifying the filtration. 


\medskip

$\bullet$ \underline{\it The SLE/CFT correspondance.}
The martingale property \eqref{eq:Z-check} is the key to establish the correspondance between SLE and CFT. Since this is not the main focus of this note -- and we already described it in several places, see e.g.~\cite{Bauer_2003,Bauer:2006growth} -- we here only present the main statement.

In the continuum, we cannot expect constructing simultaneously the measure on curves and the statistical averages, but we might expect constructing them separately and coupling them by imposing the martingale property \eqref{eq:Z-check}. On one hand,  the statistical expectation values $\vev{\Oo}^{\text{stat}}_{\DD}$ are represented in the continuum as (ratio of) statistical field theory expectation values (see below). On the other hand, we are giving a measure $\EE$ on curves $\gamma_{[0,t]}$ that are parameterized by a running variable $t$ in some way. The coupling should be such that the map $t\mapsto \vev{\Oo}^{\text{stat}}_{\DD\setminus \gamma_{[0,t]}}$ is a martingale for the $\gamma_{[0,t]}$-process. This imposes a coupling between the measure $\EE$ on curves and the field theory describing the continuum limit of the statistical model.

Assuming  conformal invariance, the measure on curves is one of the SLE measure, $\EE = \mathrm{SLE}_{\kappa}$, for some parameter $\kappa$. See \cite{Lawler:2007sle} for a review on SLE processes.

At criticality, the statistical field theory representing the continuous limit of the discrete statistical model is conformally invariant. It is thus a conformal field theory and we have:
\beq
\EE_{\DD} [\Oo] \to \vev{\Oo}_{\DD;x_0,x_\infty}^{\text{stat}} :=\frac{\vev{\Psi(x_{\infty})\,\Oo\,\Psi(x_{0})}_{\DD}^{\text{cft}}}{\vev{\Psi(x_{\infty})\,\Psi(x_{0})}_{\DD}^{\text{cft}}},
\eeq
where $\Psi(x_{0})$ and $\Psi(x_{\infty})$ are appropriate field theory observables coding for the boundary changing at $x_0$ and $x_\infty$.
Actually, this type of relation is also true away from criticality, but the peculiar property of conformal field theory is that going from $\DD\setminus \gamma_{[0,t]}$ to $\DD$ is purely kinematical, since the two domains are conformally equivalent.

The coupling then demands that $t\to \vev{\Oo}_{\DD\setminus\gamma_{[0,t]},\gamma(t),x_\infty}^{\text{stat}}$ be a SLE-martingale. In particular, for $s<t$,
\beq
\EE_\mathrm{SLE}[ \vev{\Oo}_{\DD\setminus\gamma_{[0,t]};\gamma(t),x_\infty}^{\text{stat}}\vert \gamma_{[0,s]} ]= \vev{\Oo}_{\DD\setminus\gamma_{[0,s]};\gamma(s),x_\infty}^{\text{stat}}.
\eeq
This relation relates the SLE-parameter $\kappa$ to the CFT data, since the expectation values on $\gamma_{[0,t]}$ are computed using the SLE-measure while the statistical expectation values are evaluated using the CFT. The output is that the CFT central charge is $c = (6-\kappa)(3\kappa-8)/2\kappa$ and that the boundary field $\Psi$ has conformal dimension $h = (6-\kappa) /2\kappa$. This coupling actually identifies the second order differential operator associated to the SLE process with the so-called Belavin-Polyakov-Zamolodchikov null vector operator $ \frac{\kappa}{2}L_{-1}^{2} - 2L_{-2}$ annihilating the field $\Psi$ \cite{Belavin:1984vu}.  We refer to~\cite{Bauer_2003,Bauer:2006growth} for more details.

\section{RG as an exploration process}
\label{sec:RG-exploration}

We now adapt the above framework to the Renormalization Group (RG), extracting the main overall picture at the risk of being less specific -- this global picture has of course to be detailed in each specific cases. The main probabilistic message is that RG transformations are conditioned expectation values. Depending on the reader’s taste, one may either start with the global picture presented in this Section \ref{sec:RG-exploration}, and later examine the examples in Sections \ref{sec:blocks} and \ref{sec:gaussian}, or first explore (some of) the examples to grasp the idea before turning to the global picture.
\medskip

$\bullet$ \underline{\it Exploration and filtration.} 
Imagine considering a field theory, or a statistical physics system, defined over a volume of size $L$ with a short distance cut-off $a$ (say, on a lattice with mesh $a$). Let us denote by $\mathbb{P}$ some given reference probability measure on the space $\mathcal{C}$ of field configurations and by $\mathbb{E}$ the expectation w.r.t. $\mathbb{P}$. A field theory, or a statistical model, is specified at the microscopic scale by an action $S_{\Lambda}(c)$,  depending on field configurations $c\in\mathcal{C}$, with $\Lambda:=1/a$, such that the Boltzmann weight $e^{-S_{\Lambda}}$ is (proportional to) the Randon-Nicodim derivative of the probability measure of field configurations w.r.t. to the reference measure. That is: at microscopic scale, the probability distribution of field configurations is $Z^{-1}\, e^{-S_{\Lambda}}\,\mathbb{P}$, with normalizing partition function $Z:=\mathbb{E}[e^{-S_{\Lambda}}]$.

Imagine that one is given a filtration $\mathcal{F}_\pp$ of $\sigma$-algebras over $\mathcal{C}$ such that $\mathcal{F}_\pp$ codes for information on field configurations at all length scales $(\delta x) \geq 1/\pp$. By filtration, we mean that the  $\sigma$-algebras $\mathcal{F}_\pp$ are finer as $\pp$ increases, $\mathcal{F}_\pp \subset \mathcal{F}_\qq$ for $\pp<\qq$, and as such, they specify a stochastic exploration process of the configuration space. By definition, functions measurable w.r.t. $\mathcal{F}_\pp$ allow to test field configurations at length scales larger than $1/\pp$. We shall provide examples of such filtrations in Sections \ref{sec:blocks} and \ref{sec:gaussian}.

Starting at $\pp$ small, comparable to $\pp_\mathrm{min}:= 1/L$ (i.e. $\pp_\mathrm{min}=0$ in the infinite volume limit), $\mathcal{F}_{\pp_\mathrm{min}}$ only gives access to global averaged views on field configurations. At any finite $\pp$, $\mathcal{F}_{\pp}$ allows to explore properties of field configurations at length scales larger than $1/\pp$, by testing them against $\mathcal{F}_{\pp}$-measurable functions. Increasing $\pp$ thus allows to test field configurations at smaller length scale. For $\pp$ large, comparable to $\pp_\mathrm{max}:=\Lambda$, $\mathcal{F}_{\pp_\mathrm{max}}$ provides detailed information on field fluctuations, at all available scales. Viewing the RG as a stochastic process exploring field configurations scale by scale downward the smallest slightly twists the usual formulation of the RG in the physics literature. 

Given this filtration, considering the conditional expectation values,
\beq \label{eq:conditioning}
\mathbb{E}_\pp [(\cdots)] := \mathbb{E} [(\cdots) | \mathcal{F}_\pp] ,
\eeq
amounts to integrate out all fluctuations at length scales below $1/\pp$. Integrating out small scale fluctuations is one of the usual formulation of the RG in physics.  Conditional expectation values satisfy a nested relation, $\mathbb{E}_\pp [\mathbb{E}_\qq [(\cdots)]] =\mathbb{E}_\pp [(\cdots)]$, for $\pp<\qq$, which has a simple physical interpretation: integration out  field fluctuations first at scales below $1/\qq$ and then at scales below $1/\pp$, successively, indeed amounts integrating out at once fluctuations at scales below $1/\pp$, since $1/\qq<1/\pp$.
\medskip

$\bullet$ \underline{\it The Renormalization Group.} 
Given the action $S_{\Lambda}$ at microscopic scale, one considers integrating out field fluctuations at length scales below $1/\pp$:
\beq \label{eq:def-Sp}
Z_\pp := \mathbb{E}_\pp [e^{-S_\Lambda}],\quad Z_\pp=: e^{-S_\pp}.
\eeq
The transformation $\mathcal{R}_{\pp;\Lambda}:\ S_{\Lambda}\to S_\pp$ is the Renormalization Group transformation. The action $S_\pp$, defined by $S_\pp:=-\log Z_\pp$, is called the effective action at scale $\pp$ (w.r.t. to the reference measure $\mathbb{P}$). 
By the definition \eqref{eq:def-Sp}, $e^{-S_\pp}$ is an $\EE$-martingale. Thanks to the nested structure of conditional expectation values, we have,
\beq \label{eq:partition-invariant}
Z=\mathbb{E}[e^{-S_{\Lambda}}] = \mathbb{E}[\mathbb{E}_\pp[e^{-S_{\Lambda}}]]= \mathbb{E}[e^{-S_\pp}].
\eeq
RG transformations thus preserve the partition function and amounts to organize Boltzmann sum scale by scale (similarly as discussed in Section \ref{sec:SLE} in connection with SLE).

More generally, one defines the RG transformation from $\qq$ to $\pp<\qq$ by
\beq \label{eq:def-RG}
e^{-\mathcal{R}_{\pp;\qq}(S_\qq)}:=\mathbb{E}_\pp [e^{-S_\qq}].
\eeq
The nested structure of conditional expectation values
implies that RG transformations $\mathcal{R}_{\pp;\qq}$ form a semi-group:
\beq \label{eq:semi-group}
\mathcal{R}_{\pp_1;\pp_2}\cdot \mathcal{R}_{\pp_2;\qq} = \mathcal{R}_{\pp_1;\qq},\quad \mathrm{for}\ \pp_1<\pp_2<\qq.
\eeq
In particular $\mathcal{R}_{\pp;\qq}(S_\qq)=S_\pp$ and $\mathbb{E}_\pp [e^{-S_\qq}]=e^{-S_\pp}$, for $\pp<\qq$. 

In generic cases, RG transformations $\mathcal{R}_{\pp;\qq}$ only depend on the scale separation $\tau:=\qq/\pp>1$, since changing length units corresponds to a global rescaling. We thus set  $\mathfrak{R}_\tau:=\mathcal{R}_{\pp;\tau\pp}$, or alternatively $\mathcal{R}_{\pp;\qq}=\mathfrak{R}_{\qq/\pp}$. The RG composition law then reads: $\mathfrak{R}_\tau\cdot\mathfrak{R}_{\tau'}=\mathfrak{R}_{\tau\tau'}$.

Let $\{g\}$ denote a (possibly infinite) set of variables, called coupling constants, parametrizing the actions ${S}[g]$, or alternatively the Boltzmann weights with  $w=e^{-S}$. The RG transformations mapping the action $S[g_\qq]$ into $S[g_\pp]$, for $\pp<\qq$, are then viewed as transformations of coupling constants, mapping $g_\qq$ to $g_\pp= \mathcal{R}_{\pp;\qq}(g_\qq)$. Assuming, as above, that RG transformations only depends on the scale ratio, they define a flow $g \to g_\tau:=\mathfrak{R}_\tau(g)$, generated by vector fields, called beta functions, 
\beq
\beta(g) = \tau\frac{d}{d\tau} \mathfrak{R}_\tau(g)\vert_{\tau=1} .
\eeq
Integrating the flow equation $\tau\partial_\tau g_\tau = \beta(g_\tau)$ allows to reconstruct the RG map.
Contrary to their established name, beta functions are not functions but vector fields on the space of field theory coupling constants. 

Renormalization of observables is addressed in Section \ref{sec:scaling}.


\section{Block spins}
\label{sec:blocks}


We here discuss how block spin transformations~\cite{Kadanoff1966} provide one way to construct a RG exploration process as in Section \ref{sec:RG-exploration}.
\medskip

$\bullet$ \underline{\it The 1D Ising model via RG.} 
Let us start by illustrating it with the 1D Ising model. Consider a segment of the line $\mathbb{Z}$ of length $N$, with spin variables $s_i=\pm$ on each site $i$ of this segment. The energy of a spin configuration $[s]$  is $E[s]=-J\sum_i s_is_{i+1}$ by definition (we absorb the temperature dependence in the energy scale $J$). The partition sum is $Z_N(J):=\sum_{[s]} e^{- E[s]}$. There are many ways to solve this  simple problem. We do it using RG idea, following J. Cardy's book \cite{Cardy1996}.
 
 Imagine grouping the spins by blocks of size $3$, i.e. $(\cdots][s_1s_2s_3][s_4s_5s_6][\cdots)$. Each blocks may be in $2^3=8$ configurations. We group these eight configurations in two disjoint sets to which we assign an effective spin $s'$. We may for instance choose the majority rule so that $s'=+$ if the internal spins of the block are $[++-]$ or a permutation thereof, and $s'=-$ if the internal spins are $[--+]$ up to permutation. It will  actually be simpler to assign to each block the spins of the middle site, so that the effective spin for the block $[s_1s_2s_3]$ is $s'=s_2$, or alternatively $s'=\pm$ for the configurations $[s_1\pm s_3]$. We then imagine computing the partition function in two steps: first summing over the internal spins of each blocks conditioned on their effective spins and second on the block spins. 

Consider two adjacent blocks, say $(\cdots][s_1s_2s_3][s_4s_5s_6][\cdots)$, and denote by $s'_1:=s_2$ and $s'_2:=s_5$ the two block spins. The Boltzmann weights are of the form:
\[ \cdots e^{Js_1s'_1}\times e^{Js'_1s_3}\, e^{Js_3s_4}\, e^{Js_4s'_2}\times  e^{Js'_2s_6}\cdots .\]
Doing the partial sum induces an effective interaction between the block spins. 
We sum over $s_3$ and $s_4$ at $s'_1$ and $s'_2$ fixed (the other spins $s_1,\ s_6,\cdots$ do not play a role). Using $e^{Jss'}=\cosh J(1+xss')$ with $x=\tanh J$, we may write this as the product of three terms $(\cosh J)^3(1+xs'_1s_3)(1+xs_3s_4)(1+xs_4s'_2)$. The sum over $s_3,s_4$ is done by expanding this product. It yields $2^2(\cosh J)^3\, (1+x^3\, s'_1s'_2)$. Up to a multiplicative constant (independent of the spins) this expression is of the same form as that for the original spins but with a new interaction constant $J'$,
\[ x' = x^3,\quad \mathrm{i.e.}\ \tanh J'= (\tanh J)^3.\]
The energy functional for the block spins is thus identical to the original one, $E'[s'] = N \mathfrak{e}(J) - J'\sum_i s'_is'_{i+1}$, 
up to an irrelevant constant $\mathfrak{e}(J)$, and the original partition function can be written as
\[ Z_N(J)= \sum_{[s]} e^{-E[s]} = \sum_{[s']} e^{-E'[s']} = e^{-N\mathfrak{e}(J)}\, Z_{N/3}(J').\]
There is only $N/3$ blocks if originally there was $N$ sites. We thus have effectively reduced the number of degrees of freedom from $2^N$ to $2^{N/3}$ via this blocking procedure.

By iteration, the effective coupling transforms as $x_n\to x_{n+1}=x_n^3$ at each step. There is only two fixed points: $x=1$ which corresponds to zero temperature ($J=\infty$) and $x=0$ which corresponds to infinite temperature ($J=0$). Since $x<1$, unless $J=\infty$, the effective couplings $x_n$ converge to zero, and the effective temperature increases towards infinite temperature. Hence, the long distance degrees of freedom are effectively described by an infinite temperature fixed point: they are in a disordered phase (a statement of course compatible with the absence of phase transition in 1D).

In its disordered phase, the system possesses a finite correlation length. The latter is preserved by block spin transformations, because those transformations preserve the long distance physics (they simply reshuffle the Boltzmann sum). The physical correlation length, with the dimension of a length, can be measured in units of the lattice spacing $a$. This dimensionless correlation length only depends on $J$, or equivalently on $x$. Since the lattice size has been dilated by a factor $3$ under block spin transformations, i.e. $a\to 3a$, the dimensionless correlation length $\xi$ transforms as $\xi(x') = \frac{1}{3}\xi(x)$, with $x'=x^3$. This implies that $\xi(x)=\frac{\mathrm{const.}}{\log x}= \frac{\mathrm{const.}}{\log (\tanh J)}$. It is  finite for all $x$, but it diverges exponentially close to zero temperature.
\medskip

$\bullet$ \underline{\it Block spins, conditioning and filtration.}
More generally, consider now a statistical model, defined over a finite volume of a lattice of mesh size $a$ in dimension $D$, with spin variables $s_i$ on each lattice site with energy $E$. Let us denote by $[s]$ the spin configurations and by $\mathcal{C}$ the space of spin configurations. We think about this energy functional as being the 'most general' one, with all possible interactions included, parametrized by a set of coupling constants $\{g\}$ (possibly an infinite number) of the form:
\[ E([s]; g )= \sum_{ij} g^{(2)}_{ij} s_i s_j + \sum_{ijkl} g^{(4)}_{ijkl}\,s_is_js_ks_l+\cdots.\]

Let us apply a block spin transformation~\cite{Kadanoff1966}. Each block is supposed to be of dimensionless size $b/a$ (this is its size measured in lattice units, its dimensionfull size is $b$).  For instance we can choose blocks of size $3$ in each direction, with $3^D$ sites in each block in dimension $D$. To each block we affect an effective spin $s'$, say by the majority rule or via the middle spin. To any given effective spin $s'$ corresponds a sub-set of configurations of the original spins in each block.

This blocking procedure can be iterated recursively. After $n$ steps, we get effective spin assignments $s_\delta$ for blocks of blocks etc of physical sizes $\delta=nb$. The effective spin configurations $[s_\delta]$ code for the spin configurations at length scale of order $\delta$, large compare to the microscopic scale $a$ for $n$ large enough,  and correspond to sub-sets $\mathcal{C}[s_\delta]$ of the original spin configurations. The collection of $\mathcal{C}[s_\delta]$ forms a partition of the original configuration space:
\beq
 \mathcal{C} = \bigcup_{[s_\delta]} \mathcal{C}[s_\delta] .
\eeq
Accordingly, we may decompose any sum of spin configurations in two steps, first summing over spin configurations conditioned on the block spins and then on the block spin configurations, as follows:
\beq \label{eq:partition}
\sum_{[s]\in \mathcal{C}} (\cdots) = \sum_{[s_\delta]}\sum_{[s]\in \mathcal{C}[s_\delta]} (\cdots).
\eeq
We thus have partitioned -- or conditioned -- the space of spin configurations, with the partition elements indexed by the block spin configurations $[s_\delta]$, and organized the Boltzmann sum accordingly. 

In the usual RG approach, one go upwards in length scale and looks at blocks of blocks of blocks, etc, in order to get a coarse-grained view of the spin configurations. One can alternatively read this decomposition of the configuration space in the reverse order and looks at sub-blocks of sub-blocks of sub-blocks, starting from the whole system down to the microscopic lattice mesh. By dividing blocks into sub-blocks, recursively, and associating to each sub-blocks the averaged spin value within that blocks, one progressively gains a finer small-scale view on spin configurations.

This hierarchical structure of sub-blocks yields nested partitions of the configuration space $\mathcal{C}$, and hence defines a filtration $\mathcal{F}_\pp$ of $\sigma$-algebras on $\mathcal{C}$ parametrized by the inverse of the block sizes $\pp=1/\delta$. Fonctions measurable w.r.t. to $\mathcal{F}_\pp$ are those constant on each element of the partition $\mathcal{C}[s_\delta]$ (with $\delta=1/\pp$). Consequently, $\mathcal{F}_\pp$-measurable functions provide access to information about field configurations at length scales larger than $1/\pp$.

Eq.~\eqref{eq:partition} is then identical to (\ref{eq:partition-invariant}) of Section \ref{sec:RG-exploration} if we interpret $\mathbb{E}$ as the discrete flat uniform measure on $\mathcal{C}$. We can of course modify this reference measure by factorizing the Boltzmann weights as $w_c=w^0_c w^I_c$ and use $w^0_c$ to specify the reference measure. 

The rest of the construction is as in Section \ref{sec:RG-exploration}. Eq.~\eqref {eq:def-Sp} for the partition function $Z[g] := \sum_{[s]\in \mathcal{C} } e^{-E([s]; g) }$ translates into 
\beq \label{eq:RGfondamental}
Z[g] = \sum_{[s_\delta]} Z_\delta([s_\delta];g),\quad  Z_\delta([s_\delta];g) := \sum_{[s]\in\mathcal{C}[s_\delta]}e^{-E([s];g)} 
\eeq
with $Z_\delta$ the conditioned partition function. The block spin effective energy is defined as $-\log Z_\delta([s_\delta];g) $, in a way similar to the definition of the effective action in \eqref{eq:def-Sp}. The RG hypothesis (which is here tautological as we consider the most general energy functional) is that this block spin energy is of the same nature as the original one (up to an additive constant) but with new coupling constants $g_\delta$, so that we can write 
\[ 
Z_\delta([s_\delta];g) = e^{-N \mathfrak{e}_\delta(g)}\,  e^{-E([s_\delta]; g_\delta)},
\]
The conditioned partition function $Z_\delta$ specifies the Boltzmann weights for the block spin configurations (the function $\mathfrak{e}$ does not matter as only the ratio of partition functions matter in statistical physics). 

The RG transformation is then defined as the map  $ g \to \mathfrak{R}_\delta(g):=g_\delta$.
By iteration they form a semi-group. Beta functions are the vector fields generating the RG flow. 
The dimension-full physical correlation length remains unchanged under RG transformations, because the latter only reshuffle the statistical sum. However, since the lattice mesh is rescaled from $a$ to $\delta a$, the dimensionless correlation length satisfies $\xi(g)= \delta\, \xi(g_\delta)$. See Section \ref{sec:RG-exploration}. The RG transformations cannot be iterated ad-finitum and have to be stopped once the size of the block is comparable to the correlation length, that is $a\ll \delta \ll a\xi$. The physical rational behind the renormalization group idea is that spins within blocks of size smaller than $a\xi$ behave (almost) collectively. 

In practice one can never exactly compute $ \mathfrak{R}_\delta$ as a map on an infinite set of coupling constants. The art of RG applications resides in truncating appropriately  this set keeping only the relevant coupling constants.

\section{Perturbation of Gaussian measures}
\label{sec:gaussian}

We here discuss another setup to construct a RG exploration process as in Section \ref{sec:RG-exploration} based on perturbing free field Gaussian measures.
\medskip

$\bullet$ \underline{\it Controlling length scale exploration processes.}
Let us give ourselves a family of real valued Gaussian fields $\varphi_\pp$, indexed by $\pp\in[0,\infty)$, with zero mean and covariance,
\beq
\EE[\varphi_\pp(x)\varphi_\qq(y)] = G_{\pp\wedge \qq}(x,y) ,
\eeq
where $G_\pp(x,y)$ is a chosen series of covariance defined over a domain of $\mathbb{D}\subset\mathbb{R}^D$ which, we assume, vanishes exponentially fast for $|x-y|\leq 1/\pp$. We may choose  $G_\pp= -{\Delta}^{-1} e^{\Delta/\pp^2}$ with $\Delta$ the Laplacian on  $\mathbb{D}$ with Dirichlet boundary conditions, say. 
By construction:\\
-- The field $\varphi_\infty$ is a Gaussian free field (GFF), with covariance minus the Green function of the Laplacian, while $\varphi_0$ is trivial. For $\pp$ finite, $\varphi_\pp$ has fluctuations only at length scales larger than $1/\pp$. It is smooth at very small scales.\\
-- The increments $\varphi_{\qq;\pp}(x):=\varphi_\qq-\varphi_\pp$, for $\pp<\qq$, encode fluctuations at intermediate length scales, between $1/\qq$ and $1/\pp$. In particular, $\varphi_\infty-\varphi_\pp$ codes for the fluctuations at length scales smaller that $1/\pp$.

The field $\varphi_\pp$ is a Gaussian $\EE$-martingale and can be viewed a distribution valued Brownian motion with It\^o contraction\footnote{Note that this It\^o contraction is invariant under reparaterization of $\pp$. For instance, we could chose to set $p=e^t$ and, parametrizing the field by $t$, their It\^o contraction will still be $\partial_t G_t\, dt$.},
\beq
d\varphi_\pp(x)d\varphi_\pp(y)= \dot G_\pp(x,y)\, d\pp ,
\eeq
with $\dot G_\pp:=\partial_\pp G_\pp$.  As for Brownian motion, we then introduce the filtration $\mathcal{F}_\pp$ generated by all $\varphi_\qq$ for $\qq\leq\pp$, and we define the conditional expectation value $\EE_\pp$ at scale $\pp$: 
\[
\EE_\pp[(\cdots)] := \EE[(\cdots)|\mathcal{F}_\pp] .
\]
Considering $\EE_\pp$  amounts to integrate out fluctuations at length scale smaller than $1/\pp$. 

At this stage it is clear that we made contact with the general framework of Section \ref{sec:RG-exploration}. We now slightly expand the discussion to make contact with stochastic quantization~\cite{Parisi:1981pm,Damgaard:1987vm} and with the so-called Polchinski's equation for the effective action~\cite{Polchinski:1983GV}.
\medskip

$\bullet$ \underline{\it Effective action, martingales and Polchinski's equation.}
Suppose now that the field theory is defined, at some cut-off scale $\pp_\mathrm{max}=\Lambda$, as a perturbation of this Gaussian measure with Boltzmann weights $e^{-V_\Lambda}$, so that the (interacting) field theory measure is
\beq \label{eq:interac-measure}
\hat{\mathbb{E}}^\Lambda[(\cdots)] := \mathbb{E}[ e^{-V_\Lambda(\varphi_\Lambda)}(\cdots)],
\eeq
with a potential $V_\Lambda$ normalized such that $\EE[e^{-V_\Lambda(\varphi_\Lambda)} ]=1$. We view $a:=1/\Lambda$ as a short distance cut-off analog of the lattice mesh of previous Section \ref{sec:blocks}.  Introducing the cut-off $\Lambda$ ensures the absence of short distance singularity.  

The nested property of conditional expectation values implies\footnote{Since $\varphi_\pp$ is smooth at short distances, there is no constraint in using non-linear  functions (no short distance singularities).}, for any $\mathcal{F}_\pp$-measurable function $F(\varphi_\pp)$, 
\beq \label{eq:twistingGFF}
\hat{\EE}^\Lambda[F(\varphi_\pp)]=  \EE[Z_\pp F(\varphi_\pp)] ~,
\eeq
with $Z_\pp:=\EE_\pp[e^{-V_\Lambda(\varphi_\Lambda)}]$, the (relative) partition function at scale $\pp$. By construction,  $Z_\pp$ is an  $\EE$-martingale, and Eq.~\eqref{eq:twistingGFF} represents the interacting measure (up to scale $\pp$) as a twisting of the Gaussian measure by a martingale. 

According to Section \ref{sec:RG-exploration}, the effective action at scale $\pp$ is defined as $S_\pp :=-\log Z_p$. Since the field increments $\varphi_\Lambda-\varphi_\pp$ are $\EE$-independent of $\varphi_\pp$, we have $Z_\pp=z_\pp(\varphi_\pp)$, with 
\beq
z_\pp(\varphi) := \EE[\, e^{-V_\Lambda(\varphi+\varphi_\Lambda-\varphi_\pp)}\,],
\eeq
and the effective action can alternatively be represented as
\beq \label{eq:Z-Seffective}
S_\pp(\varphi) := - \log z_\pp(\varphi) ~.
\eeq
As a conditional expectation value, $Z_\pp$ is an $\EE$-martingale, by construction.  As a consequence, the drift term of its It\^o derivative vanishes, so that,
\beq
\partial_\pp z_\pp(\varphi) + \frac{1}{2} \int\!\! dxdy\, \dot G_\pp(x,y)\, \delta_x\delta_y z_\pp(\varphi)=0 ,
\eeq
where we set $\delta_x := \frac{\delta}{\delta\varphi(x)}$. Alternatively,  
\beq \label{eq:polchinski-martingale}
\partial_\pp S_\pp(\varphi) = \frac{1}{2} \int\!\! dxdy\, \dot G_\pp(x,y) \Big(  \delta_xS_\pp(\varphi)\delta_y S_\pp(\varphi) - \delta_x\delta_y S_\pp(\varphi) \Big) .
\eeq
This equation is known as Polchinski's equation \cite{Polchinski:1983GV}. It is a flow equation over length scales, with boundary conditions at the cut-off scale: $S_\Lambda=V_\Lambda$. It encodes for numerous, if not all, information about the field theory. It is  however difficult, if not impossible, to solve in general, except in the Gaussian case. Deriving Polchinski's equation from the martingale property has recently been discussed in~\cite{Bauerschmidt_2024}.

The representation \eqref{eq:twistingGFF} allows to make contact with stochastic quantization~\cite{Parisi:1981pm,Damgaard:1987vm}. Since $dZ_\pp=-Z_\pp\, \delta S_\pp(\varphi_\pp)\, d\varphi_\pp$, Girsanov's theorem~\cite{Oksendal-book} guarantees that $\varphi_\pp$ is solution of a stochastic differential equation, whose drift is parametrized by the effective action, 
\beq \label{eq:SDE-stoQ}
d\varphi_\pp = - \dot G_\pp\cdot \delta S_\pp(\varphi_\pp)\, d\pp + d\hat\varphi_\pp,
\eeq
with $\pp$ as effective time parameter and with $\hat\varphi_\pp$ an $\hat{\EE}^\Lambda$-Brownian motion with It\^o contraction $d\hat \varphi_\pp(x)d\hat \varphi_\pp(y)= \dot G_\pp(x,y)\, d\pp$. We leave as an exercice to the reader to verify that $ \delta S_\pp=\hat{\EE}^\Lambda_\pp[\delta V_\Lambda(\varphi_\Lambda)]$ so that the drift in \eqref{eq:SDE-stoQ} is an $\hat{\EE}^\Lambda$-martingale. As a consequence, the law of $\varphi_\Lambda$ solution of \eqref{eq:SDE-stoQ} 
is $e^{-V_\Lambda(\varphi_\Lambda)}d\mu_{G_\Lambda}(\varphi_\Lambda)$, with $\mu_{G_\Lambda}$ the law of the GFF with variance $G_\Lambda$. Eq.~\eqref{eq:SDE-stoQ} thus enables the transport of the Gaussian measure into the interacting measure~\eqref{eq:interac-measure}. Using the length scale $\pp$ as an effective time parameter introduces a subtle variation from the conventional approach to stochastic quantization. This viewpoint has recently been advocated in~\cite{bailleul2023wilsonitodiffusions}. 

\section{Scaling limit and renormalization}
\label{sec:scaling}

Constructing field theories in the continuum requires taking the limit of vanishing short distance cut-off. Let us imagine giving ourselves a lattice microscopic statistical model\footnote{If instead one prefers to think about the field theory as a perturbation of a Gaussian theory, the short distance cut-off $a$ is the inverse of the large regularizing parameter $\Lambda$ introduced in Section  \ref{sec:gaussian}.}, with lattice mesh $a$ as in Section \ref{sec:blocks}, to have a definite picture in mind. We then wonder how to take the limit $a\to 0$ to properly define a field theory in the continuum. Our discussion will be based on physically expected hypothesis and aims to enlighten the global picture. It requires a detailed, case by case, analysis to become mathematically rigorous.
\medskip

$\bullet$ \underline{\it Scaling and continuous limits.} Geometrically, the continuous limit is the limit of vanishing lattice mesh $a\to 0$. If some points $x_k$ are marked, say indicating the insertion of local observables, the limit $a\to 0$ goes together with the limit ${\tt n}_k\to \infty$ with $x_k=a{\tt n}_k$ fixed, if ${\tt n}_k$ denote the integer lattice coordinates of the marked points. 

The general principle underlying the construction of field theories in the continuum is that this geometrical limit must be paired with a proper adjustment of the coupling constants. This combined limit is called the scaling limit. If $\{g_a\}$ denotes the set of coupling constants of the microscopic lattice model, their dependence on the mesh size $a$ is fixed by the RG transformations. Let $\ell_R$ and $g_R$ be respectively arbitrarily chosen length and coupling constants, the scaling limit is then defined as
\beq
\label{eq:scaling-lim}
\lim_{\mathrm{scaling}} := \lim\Big( a\to0\ \mathrm{s.t.}\ \ell_R\ \&\ \mathfrak{R}_{\ell_R/a}(g_a) = g_R\  \mathrm{fixed}\Big),
\eeq
with $\mathfrak{R}_{\tau}$ the RG flow map as in Section \ref{sec:RG-exploration}.
In physics literature, $g_R$ and $\ell_R$ are respectively named the renormalized coupling constants and the renormalization scale. Their choice is arbitrary but, as we will see below, the final theory is covariant under changing them.

Eq.~\eqref{eq:scaling-lim} fixes how $g_a$ is adjusted when $a\to 0$. It requires implementing (and thus controlling) the RG flow a long way as $\ell_R/a\to\infty$ as $a\to 0$. Let us analyze it in the neighborhood a RG fixed point $g^*$, such that $\mathfrak{R}_\tau(g^*)=g^*$. Near $g^*$, some directions in the space of coupling constants are attractive, some are repulsive -- and some are marginal, but for simplicity we do not consider this possibility here. The repulsive directions are called relevant, the attractive ones irrelevant. The latter specify  hyper-surface called the critical hyper-surface.  Irrelevant coupling constants flow to zero along the RG flow. A theory is said to be renormalizable if there is only a finite number of relevant directions, or alternatively, if the critical hyper-surface has finite co-dimension. Eq.~\eqref{eq:scaling-lim}  imposes to the coupling constants $g_a$ to approach the critical hyper-surface as $a\to 0$. 

For illustration purposes, let us suppose that there is only a single relevant direction. Let $g$ be the relevant coupling constant, with a fixed point at $g^*=0$. Let us assume\footnote{up to redefining the coupling constant or for $g\ll1$.} that the beta function is linear $\beta(g)= y g$, with $y>0$ for $g$ to be relevant, so that $\mathfrak{R}_\tau(g)= \tau^yg$. Eq.~\eqref{eq:scaling-lim} reads $g_R/g_a=(\ell_R/a)^y$, or alternatively $g_a= (a/\ell_R)^y g_R$. This indeed specifies how $g_a$ approaches the fixed point as $a\to0$. By a similar argument as in the Ising case, see Section \ref{sec:blocks}, the dimensionless correlation length satisfies $\xi(\mathfrak{R}_\tau(g))=\xi(g)/\tau$, so that $\xi(g_a)\simeq |g_a|^{-1/y}$, for $g_a$ near the fixed point. The physical dimension-full correlation length thus behaves as $\xi_\mathrm{phys}\simeq a|g_a|^{-1/y}= \ell_R |g_R|^{-1/y}$, or alternatively $g_R=(\ell_R/\xi_\mathrm{phys})^y$. 

In other words, the scaling relation $\mathfrak{R}_{\ell_R/a}(g_a) = g_R$ ensures the finiteness of the physical dimension-full correlation length in the scaling limit, as required for a meaning-full continuous theory.
\medskip

$\bullet$ \underline{\it Renormalized observables.}
To construct correlation functions in the continuum we first need to analyze how observables transform under the RG. This construction is analogous to that done in Section \ref{sec:RG-exploration} for the partition functions, and we use the notations of Section \ref{sec:RG-exploration}. Recall that an observable $\mathcal{O}$ is a function on the configuration space: $c\in\mathcal{C}\to \mathcal{O}_c$. At microscopic scale, with mesh $a=1/\Lambda$, $\mathcal{O}$ is $\mathcal{F}_\Lambda$-measurable. If, as in Section \ref{sec:RG-exploration}, the microscopic probability distribution of field configurations is $\hat{\mathbb{P}}:=Z^{-1}\, e^{-S_{\Lambda}}\,\mathbb{P}$, with microscopic action $S_\Lambda$ and $Z=\mathbb{E}[e^{-S_{\Lambda}}]$, then expectation values of observables are given by:
\beq \label{eq:O-expect-Lambda}
\hat \EE[\mathcal{O}]=Z^{-1}\, \EE[e^{-S_\Lambda}\mathcal{O}].
\eeq
As for partition functions, we compute them by first conditioning on field configurations at length scale larger than $1/\pp$, i.e. by conditioning on $\mathcal{F}_\pp$, so that $\hat \EE[\mathcal{O}]= \hat \EE[\mathcal{O}_\pp]$ with $\mathcal{O}_\pp:=\hat \EE[\mathcal{O}|\mathcal{F}_\pp]$. By the rules of conditional expectation values, we have 
\beq
\mathcal{O}_\pp:=\hat \EE[\mathcal{O}|\mathcal{F}_\pp]=\frac{\EE_\pp[e^{-S_\Lambda}\mathcal{O}]}{\EE_\pp[e^{-S_\Lambda}]} = Z_\pp^{-1}\, \EE_\pp[e^{-S_\Lambda}\mathcal{O}],
\eeq
with $Z_\pp= \EE_\pp[e^{-S_\Lambda}]$, as in \eqref{eq:def-Sp}. By construction, $\mathcal{O}_\pp$ are $\hat \EE$-martingales and, in particular, $\mathcal{F}_\pp$-measurables.

As a consequence, we write
\beq \label{eq:O-expect-pp}
\hat \EE[\mathcal{O}]=\hat \EE[\mathcal{O}_\pp]=Z^{-1}\,\EE[e^{-S_\pp}\mathcal{O}_\pp],
\eeq
with $S_\pp=-\log Z_\pp$ the effective action and $Z=\EE[e^{-S_\pp}]$. 
Let us compare \eqref{eq:O-expect-Lambda} and \eqref{eq:O-expect-pp}: Starting from an observable $\mathcal{O}$ defined at the microscopic scale $1/\Lambda$, with Boltzmann weights $e^{-S_\Lambda}$ at that scale, we transformed it into an observable defined at a larger length scale $\ell=1/\pp$ with Boltzmann weights $e^{-S_\pp}$ specified by the effective action at that scale. If at scale $a=1/\Lambda$, the action $S_\Lambda=S[g_a]$ is parametrized by the set of coupling constants $\{g_a\}$, then at scale $\ell=1/\pp$ the effective action $S_\pp=S[g_\pp]$ is parametrized by the coupling constant $g_\pp=\mathfrak{R}_{\ell/a}(g_a)$, computed using the RG flow as in Section \ref{sec:RG-exploration}.

More generally, we define the RG transformation of observables from $\qq$ to $\pp<\qq$ by
\beq \label{eq:R-martingale}
\hat{\mathcal{R}}^{[g_\qq]}_{\pp;\qq}(\mathcal{O}_\qq) :=\hat \EE[ \mathcal{O}_\qq | \mathcal{F}_\pp] = e^{+S_\pp}\, \EE_\pp[e^{-S_\qq}\mathcal{O}_\qq],
\eeq
with $\EE_\pp[e^{-S_\qq}]=e^{-S_\pp}$.
RG transformation of observables differs from the renormalization of the action in that the former is defined as conditional expectation values w.r.t. to $\hat \EE$ while the latter is defined as conditional expectation values w.r.t. to $\EE$. It depends on the coupling constant $g_\qq$ since $S_\qq=S[g_\qq]$.
Their composition law reads
\beq
\hat{\mathcal{R}}^{[g_{\pp_2}]}_{\pp_1;\pp_2}\cdot \hat{\mathcal{R}}^{[g_\qq]}_{\pp_2;\qq}=\hat{\mathcal{R}}^{[g_\qq]}_{\pp_1;\qq},\quad \mathrm{for}\ \pp_1<\pp_2<\qq.
\eeq
We may, and will, assume that it only depends on the ratio of the scale: $\hat{\mathcal{R}}^{[g]}_{\pp;\qq}=\hat{\mathfrak{R}}^{[g]}_{\qq/\pp}$. The composition law then reads: 
\beq 
\label{eq:RG-coycle-big}
\hat{\mathfrak{R}}^{[\mathfrak{R}_\delta(g)]}_{\delta'}\cdot \hat{\mathfrak{R}}^{[g]}_{\delta}= \hat{\mathfrak{R}}^{[g]}_{\delta'\delta} .
\eeq

The martingale property \eqref{eq:R-martingale} ensures that:
\beq \label{eq:O-invariance}
Z^{-1}\, \EE[ e^{-S[g_\pp]}\, \hat{\mathfrak{R}}^{[g_\qq]}_{\qq/\pp}(\mathcal{O}_\qq)] = Z^{-1}\,\EE[e^{-S[g_\qq]}\, \mathcal{O}_\qq],
\eeq
with $g_\pp={\mathfrak{R}}_{\qq/\pp}(g_\qq)$, for $\pp<\qq$. 
As Eq.~\eqref{eq:O-expect-pp}, this equation codes for the invariance of expectation values under the RG flow: The expectation values of $\mathcal{O}_\qq$ with Boltzmann weights with coupling constants $g_\qq$ is equal to that of its RG transformed $\hat{\mathfrak{R}}^{[g_\qq]}_{\qq/\pp}(\mathcal{O}_\qq)$ with Boltzmann weights with coupling constants $g_\pp$.

Suppose now that the observable is microscopically defined as a product of local observables, $\mathcal{O} = \prod_k \Phi_k({\bf n}_k)$, where each $\Phi_k$ is an observable sensitive to the local environment near its insertion lattice point (say ${\tt n}_k$, with ${\tt n}$ the dimensionless distance counted in unit of lattice spacing). Let us assume that we are giving a complete list of local observables -- of course, at this discussion level this assumption is formal and requires a better control to make it mathematically rigorous.
It is physically reasonable to expect that block spin transformations map local observables into local observables -- because spins behave collectively within blocks of sizes smaller than the correlation length, see Section \ref{sec:blocks}. It is thus physically reasonable to assume that local observables are renormalized into local observables, so that we can write (hiding field labels to simplify the notation):
\beq \label{eq:Gamma-anomalous}
\hat{\mathfrak{R}}_\delta^{[g]}(\Phi({\tt n}) ) = \Gamma_\delta^{[g]}\cdot \Phi({\tt n}/\delta),
\eeq
where $\Gamma_\delta^{[g]}$ is a matrix, coding for the decomposition of the transformed observable on the basis of the local observables. The rescaling of the lattice position of the operators comes about by noticing that, after a block spin transformation, the dimensionless distances, counted by the number of lattice spacing, has been divided by $\delta$. 

The matrices $\Gamma_\delta^{[g]}$, which depend both on the rescaling factor $\delta$ and on the coupling constants, are called the mixing matrices or the matrices of anomalous dimensions, equivalently. They reflect how local observables are mixed under RG transformations. From the RG composition law \eqref{eq:RG-coycle-big}, they inherit a cocycle structure:
\beq \label{eq:G-cocyle}
 \Gamma_{\delta}^{[g]}\cdot \Gamma_{\delta'}^{[\mathfrak{R}_\delta(g)]} = \Gamma_{\delta\delta'}^{[g]}.
 \eeq

At a RG fixed point $g^*$, they transform multiplicatively and commutatively: $\Gamma_{\delta'}^{[g^*]}\cdot \Gamma_{\delta}^{[g^*]} = \Gamma_{\delta\delta'}^{[g^*]}$. The local observables $\Phi_i$ with proper scaling dimension are those diagonalizing them: $\Gamma_{\delta}^{[g^*]}\cdot \Phi_i = \delta^{-\Delta_i}\, \Phi_i$, with $\Delta_i$ the (conformal) scaling dimension of $\Phi_i$. 

The RG invariance of expectation values \eqref{eq:O-invariance} translates into:
\beq \label{eq:E-Phi-invariance}
Z^{-1}\, \EE[ e^{-S[\mathfrak{R}_\delta(g_a)]}\, \prod_k\Gamma_\delta^{[g]}\cdot \Phi_k({\tt n}_k/\delta)] = Z^{-1}\,\EE[e^{-S[g_a]}\, \prod_k\Phi_k({\tt n}_k)],
\eeq 
with $a=1/\Lambda$ the lattice mesh and denoting $\hat{\mathfrak{R}}_\delta^{[g]}(\Phi_k({\tt n}_k) )$ by $\Gamma_\delta^{[g]}\cdot \Phi_k({\tt n}_k/\delta)$ as in \eqref{eq:Gamma-anomalous}.
\medskip

$\bullet$ \underline{\it Scaling limit of fields and renormalized correlation functions.}
At a RG fixed point, local fields\footnote{In this last Section, we use the label "cont." and "latt." to distinguished observables respectively defined in the continuum or on the lattice.} in the continuum are defined by dressing lattice observables in an $a$-dependent way according to their anomalous dimension: $\Phi_i^\mathrm{cont.}(x) = \lim_{a\to 0} {a}^{-\Delta_i}\, \Phi^\mathrm{latt.}_i({\tt n}={x}/{a})$, with $\Delta_i$ the scaling dimension. Away from RG fixed points, this dressing has to be paired with the running of the coupling constants $g_a$ according to the RG.  The proper definition uses the matrix of anomalous dimension \eqref{eq:Gamma-anomalous}:
\beq \label{eq:def-phi-continuum}
 {\Phi}^\mathrm{cont.}(x):= \lim_{\mathrm{scaling}} \hat \Gamma^{-1}_a\cdot{\Phi}^\mathrm{latt.}({\tt n}={x}/{a}), 
 \quad \hat \Gamma_a := \Gamma_{\ell_R/a}^{[g_a]} ,
 \eeq
where the scaling limit is defined in \eqref{eq:scaling-lim} with $\ell_R$ and $g_R$ the (chosen) renormalization scale and coupling constants, respectively. Here $\Gamma_{\ell_R/a}^{[g_a]}$ is the matrix of anomalous dimensions defined in \eqref{eq:Gamma-anomalous}. Note that $g_a$ is a function of $g_R$ and $\ell_R/a$ via \eqref{eq:scaling-lim}.

The definition \eqref{eq:def-phi-continuum} ensures that the scaling limit of the expectation values of products of local observables $\Phi_k$, dressed by the mixing matrices, (formally) exists. Indeed, imagine changing slightly the lattice mesh $a\to \lambda a$, with $|\log\lambda|\ll1$, and interpret such rescaling as a RG transformation. The  cocyle relation \eqref{eq:G-cocyle} with $\delta=\lambda$ and $\lambda \delta'=\ell_R/a$ yields $\hat \Gamma_{\lambda a}^{-1}= \hat \Gamma_a^{-1}\, \Gamma_\lambda^{[g_a]}$. The RG invariance of expectation values \eqref{eq:O-invariance} then (formally) implies that the following expectation values are $a$-independent: 
\beq
Z^{-1}\, \EE[ e^{-S[g_a] }\, \prod_k\hat \Gamma_a^{-1}\cdot {\Phi}_k^\mathrm{latt.}({\tt n_k}={x_k}/{a}) ]  .
\eeq
Of course, making this argument mathematically rigorous requires controlling the RG transformations on coupling constants and on observables. A possible way consists in truncating the RG to a finite set of coupling constants and local observables at the cost of making this independence only  approximate and asymptotic at small cut-off $a$ and controlling the reminders for small $a$.

Although formal, this property indicates (at this level of precision) that the proper definition of the renormalized expectation values, in the continuum, is:
\beq \label{eq:def-continuum}
\langle \prod_k {\Phi}_k^\mathrm{cont.}(x_k) \rangle^\mathrm{R}_{\ell_R,g_R} := \lim_{\substack{ \mathrm{scaling} \\ x_k=a{\tt n}_k\ \mathrm{fixed}}} Z^{-1}\, \EE[ e^{-S[g_a] }\, \prod_k(\Gamma_{\ell_R/a}^{[g_a]} )^{-1}\cdot {\Phi}_k^\mathrm{latt.}({\tt n_k}) ] .
\eeq
This definition involves the RG flow, via the relation $\mathfrak{R}_{\ell_R/a}(g_a)=g_R$, and the RG dressing of local observables, via the mixing matrix $\Gamma_{\ell_R/a}^{[g_a]}$. Constructing field theories in the continuum, and their renormalizability, amounts to prove that the limit \eqref{eq:def-continuum} makes sense.

By construction, the renormalized expectation values \eqref{eq:def-continuum} depend on the chosen renormalization scale $\ell_R$ and coupling constants $g_R$. In other words, the renormalized expectation values depend on which scale $\ell_R$ the renormalized coupling constants $g_R$ are specified. However, this dependency is RG covariant, in the sense that the renormalized expectation values \eqref{eq:def-continuum} are invariant along the RG flow:
\beq \label{eq:callan-symanzik}
\langle \prod_k\hat{\mathfrak{R}}^{[g_R]}_\lambda\big({\Phi}^\mathrm{cont.}_k(\lambda {x_k})\big) \rangle^R_{\lambda \ell_R, \mathfrak{R}_\lambda(g_R)} 
= \langle \prod_k {\Phi}^\mathrm{cont.}_k(x_k) \rangle^R_{\ell_R,g_R},
\eeq 
for any rescaling $\lambda$, with $\hat{\mathfrak{R}}^{[g_R]}_\lambda\big({\Phi}^\mathrm{cont.}_k(\lambda x_k)\big):=\Gamma^{[g_R]}_\lambda\cdot{\Phi}^\mathrm{cont.}_k({x_k})$. In its infinitesimal form (i.e. for $|\log \lambda|\ll1$), this covariance property  is called the Callan-Symanzik equation~\cite{Callan1970,Symanzik1970}. It is important as it allows to decipher the short and large distance properties of renormalized expectation values.

A (maybe not so usual) proof of \eqref{eq:callan-symanzik} uses the RG  composition law \eqref{eq:semi-group} and cocycle relation \eqref{eq:G-cocyle}. Since $\mathfrak{R}_\lambda(g_R)= \mathfrak{R}_{\lambda\ell_R/a}(g_a)$ from \eqref{eq:semi-group}, transforming $\ell_R,\,g_R$ into $\lambda\ell_R,\,\mathfrak{R}_\lambda(g_R)$ leaves $g_a$ invariant. By definition \eqref{eq:def-continuum}, we then have: 
 \beq \label{eq:proof-CS}
 \langle \prod_k \Gamma_\lambda^{[g_R]}\cdot \Phi^\mathrm{cont.}_k(x_k)\rangle^R_{\lambda \ell_R,\mathfrak{R}_\lambda(g_R)} = \!\!\!\lim_{\substack{ \mathrm{scaling} \\ x_k=a{\tt n}_k\ \mathrm{fixed}}} \!\!\! Z^{-1}\, \EE[ e^{-S[g_a] }\, \prod_k \Gamma_\lambda^{[g_R]} (\Gamma_{\lambda\ell_R/a}^{[g_a]} )^{-1}\cdot {\Phi}_k^\mathrm{latt.}({\tt n_k}) ]
 \eeq
From the cocycle relation \eqref{eq:G-cocyle}, with $\delta'=\lambda$, $\delta=\ell_R/a$, and  $g=g_a$ so that $\mathfrak{R}_\delta(g_a)=g_R$, we have:
\beq \label{eq:cocycle-bis}
 \Gamma_{\ell_R/a}^{[g_a]}\, \Gamma_{\lambda}^{[g_R]} = \Gamma_{\lambda\ell_R/a}^{[g_a]}.
 \eeq
Hence, $\Gamma_\lambda^{[g_R]} (\Gamma_{\lambda\ell_R/a}^{[g_a]} )^{-1}=(\Gamma_{\ell_R/a}^{[g_a]} )^{-1}$, and the scaling limit of r.h.s. of \eqref{eq:proof-CS} is  effectively $\langle \prod_k {\Phi}_k(x_k) \rangle^R_{\ell_R,g_R}$. This proves (for physics standard) the RG covariance \eqref{eq:callan-symanzik}. Of course a mathematically complete proof requires controlling enough the RG transformations and the mixing matrices. 
 
 To conclude: controlling -- when possible -- RG transformations of coupling constants and fields allows to define field theory in the continuum according to \eqref{eq:def-continuum}. We then automatically get the Callan-Symanzik covariance property as a byproduct of RG composition laws \eqref{eq:G-cocyle}.

\section{Conclusion and perspective}
\label{sec:conclusion}

The two processes we discussed -- the Schramm-Loewner Evolution for conformal interfaces and the Renormalization Group -- illustrate how phase space exploration processes are powerful tools to understand random geometrical or statistical physics models, and their scaling limits. Framing these explorations through a filtration of progressively finer $\sigma$-algebras, each encoding increasingly detailed information about field configurations, provides a synthetic and unifying perspective.

Let us conclude by mentioning two open questions worth addressing. First, given that many physically relevant models exhibit distinct effective degrees of freedom at short (UV) and long (IR) scales, can the RG exploration processes be generalized to scenarios where the nature of these degrees of freedom evolves along the flow? Second, given the unified perspective provided by exploration processes, beyond SLE and RG, are there other physically meaningful, tractable exploration processes for probing field configuration spaces?

\bigskip

{\it Acknowledgements}: We thank Michel Bauer for our past collaboration on SLE processes (around 2002-2006), during which this approach to the RG as an exploration process emerges. We also thank Xin Sun for the organization of ''Peking summer program in probability: Probability \& Quantum Field Theory", June 1-19, 2026, in Peking (China), where part of this material was presented, and for encouraging me to write these notes. DB is partly supported by the CNRS, the ENS and the Simons Foundation via the Simons Collaboration on Probabilistic Paths to QFT.


\bibliography{RG-exploring}{}
\bibliographystyle{plain}


\end{document}